\documentclass{ieeeaccess}
\usepackage{cite}
\usepackage{amsmath,amssymb,amsfonts}
\usepackage{algorithmic}
\usepackage{graphicx}
\usepackage{textcomp}
\usepackage{lipsum}

\usepackage[english, russian]{babel}
\usepackage[T1,X2]{fontenc}
\usepackage[utf8]{inputenc}

\usepackage{xcolor}
\usepackage[normalem]{ulem}
\usepackage{hyperref}
\hypersetup{
    colorlinks = true,
    linkbordercolor = {},
    allcolors  = blue
}

\def\BibTeX{{\rm B\kern-.05em{\sc i\kern-.025em b}\kern-.08em
    T\kern-.1667em\lower.7ex\hbox{E}\kern-.125emX}}
\begin{document}

\history{Date of publication xxxx 00, 0000, date of current version xxxx 00, 0000.}
\doi{10.1109/ACCESS.2017.DOI}

\title{{
Nonlinear State Space Modeling and Control of the Impact of Patients' Modifiable Lifestyle Behaviors on the Emergence of Multiple Chronic Conditions}
}
\author{\uppercase{Syed Hasib Akhter~Faruqui}\authorrefmark{1}, 
\uppercase{Adel~Alaeddini\authorrefmark{1}, Jing~Wang\authorrefmark{2}
Susan~P~Fisher-Hoch\authorrefmark{3}, and Joseph~B~Mccormick\authorrefmark{3
}
},
}
\address[1]{Department of Mechanical Engineering, The University of Texas at San Antonio, San Antonio, TX 78249. (e-mail:  \{syed-hasib-akhter.faruqui, adel.alaeddini\}@utsa.edu)}
\address[2]{School of Nursing, UT Health San Antonio, San Antonio, TX 78229.
}
\address[3]{School of Public Health Brownsville, The University of Texas Health Science Center at Houston, TX 78520.
}

\tfootnote{This research work was supported by the National Institute of General Medical Sciences of the National Institutes of Health under award number "1SC2GM118266-01".}

\markboth
{Author \headeretal: Preparation of Papers for IEEE TRANSACTIONS and JOURNALS}
{Author \headeretal: Preparation of Papers for IEEE TRANSACTIONS and JOURNALS}

\corresp{Corresponding author: Adel Alaeddini (e-mail: \url{adel.alaeddini@utsa.edu}).}

\begin{abstract}\\
The emergence and progression of multiple chronic conditions (MCC) over time often form a dynamic network that depends on patients' modifiable risk factors and their interaction with non-modifiable risk factors and existing conditions.
Continuous time Bayesian networks (CTBNs) are effective methods for modeling the complex network of MCC relationships over time. However, CTBNs are not able to effectively formulate the dynamic impact of patient's modifiable risk factors on the emergence and progression of MCC. Considering a functional CTBN (FCTBN) to represent the underlying structure of the MCC relationships with respect to individuals' risk factors and existing conditions, we propose a nonlinear state-space model based on Extended Kalman filter (EKF) to capture the dynamics of the patients' modifiable risk factors and existing conditions on the  MCC evolution over time. We also develop a tensor control chart to dynamically monitor the effect of changes in the modifiable risk factors of individual patients on the risk of new chronic conditions emergence.
We validate the proposed approach based on a combination of simulation and real data from a dataset of 385 patients from Cameron County Hispanic Cohort (CCHC) over multiple years. The dataset examines the emergence of 5 chronic conditions (Diabetes, Obesity, Cognitive Impairment, Hyperlipidemia, and Hypertension) based on 4 modifiable risk factors representing lifestyle behaviors (Diet, Exercise, Smoking Habit, and Drinking Habit) and 3 non-modifiable risk factors, including demographic information (Age, Gender, Education). The results demonstrate the effectiveness of the proposed methodology for dynamic prediction and monitoring of the risk of MCC emergence in individual patients.

\end{abstract}
\titlepgskip=-15pt

\maketitle
\section{Introduction}
\label{Section: Introduction}
\IEEEPARstart{T}{he} evolution of multiple chronic conditions (MCC) follows a complex stochastic process. This path of evolution is often influenced by several factors, including inter-relationship of existing conditions, patient-level modifiable and non-modifiable risk factors \cite{rappaport_genetic_2016}. 
MCCs are associated with 66\% of the total healthcare costs in the United States, and approximately one in four Americans and 75\% of Americans aged 65 years are burdened with MCC \cite{goodman_defining_2013, campbell_reducing_2017}. 
Furthermore, people with MCCs have an increased risk of mortality \cite{prados-torres_multimorbidity_2014}. Thus having MCC is one of the biggest challenges of the 21st century in healthcare \cite{baker_crossing_2001}. 

Several aspects of MCC have been studied in literature over the years. Lippa et al. \cite{lippa_deployment-related_2015} conducted a structured clinical interview of a sample of 255 previously deployed Post-9/11 service members and veterans. They found over 90\% of them suffer from psychiatric conditions. Approximately half of them had three or more conditions, and 76.9\% of them suffer from four clinically relevant psychiatric and behavioral factors, including deployment trauma, somatic, anxiety, and substance abuse. Alaeddini et al. \cite{alaeddini_mining_2017} identified major transitions of four MCC that include hypertension (HTN), depression, PTSD, and back pain in a cohort of 601,805 Iraq and Afghanistan war Veterans (IAVs). They also developed a Latent Regression Markov Mixture Clustering (LRMCL) algorithm that can predict the exact status of comorbidities about 48\% of the time. In a separate study, Cai et al. \cite{cai_analysis_2015} developed algorithms to identify the relationships between factors influencing hepatocellular carcinoma after hepatectomy. Lappenschaar et al.\cite{lappenschaar_multilevel_2013} and Faruqui et al.\cite{faruqui_mining_2018} separately used a large dataset to develop a multilevel temporal Bayesian network (MTBN) to model the progression of MCCs. Several studies have also covered the prevalence of MCC and their rate of increase \cite{wolff_prevalence_2002, vogeli_multiple_2007, anderson_growing_2004, schneider_prevalence_2009, freid_multiple_2012, lehnert_review_2011, ward_prevalence_2013, lochner_county-level_2015, ward_multiple_2014, cabassa_race_2013}; health consequences of MCC and their complications \cite{bayliss_predicting_2004, tinetti_contribution_2011, grembowski_conceptual_2014, hempstead_fragmentation_2014, gijsen_causes_2001}; cost and quality of life \cite{friedman_hospital_2006, chen_health-related_2010, boyd_clinical_2005, fried_health_2011, zhong_effect_2015, min_multimorbidity_2007, domino_heterogeneity_2014}; patient support, intervention and complications \cite{wyatt_out_2014, beadles_medical_2015}; and assessment, prediction, and decision making \cite{pugh_complex_2014, miotto_deep_2016, alaeddini_mining_2017, faruqui_mining_2018}. However, most of the existing literature is cross-sectional, considers single chronic conditions, or studies a short period of time. Moreover, while these methods describe general comorbidity phenotypes, they do not provide insight into 
the impact of existing comorbid conditions, and lifestyle behaviors of individual patients on dynamics of MCC emergence and progression \cite{faruqui_mining_2018}. Therefore, the dynamic interactions between MCCs and lifestyle behaviors of an individual patient on the complex evolution pathway of
MCC are not precisely known and need further investigation.

In this study, we first represent the complex stochastic relationship between MCC as a functional continuous time Bayesian network (FCTBN) \cite{faruqui_functional_2020} to take into account the impact of the patients' risk factors on the MCC emergence and progression. We then develop a dynamic FCTBN (D-FCTBN) to capture the dynamic impact of modifiable risk factors and their interaction with existing conditions on the emergence of new MCC. This is done by formulating the conditional dependencies of FCTBN using a non-linear state-space model based on Extended Kalman Filter (EKF). Next, We develop a tensor control chart to monitor the changes in the estimated parameters of the proposed D-FCTBN model, which may have a potentially significant impact on the risk of developing a new MCC. Finally, we validate the proposed approach using a combination of simulation and real data. The overall schema for the proposed method is shown in Figure \ref{Figure:Overall_Planning}.

\Figure[t!](topskip=0pt, botskip=0pt, midskip=0pt)[width=.90\textwidth]{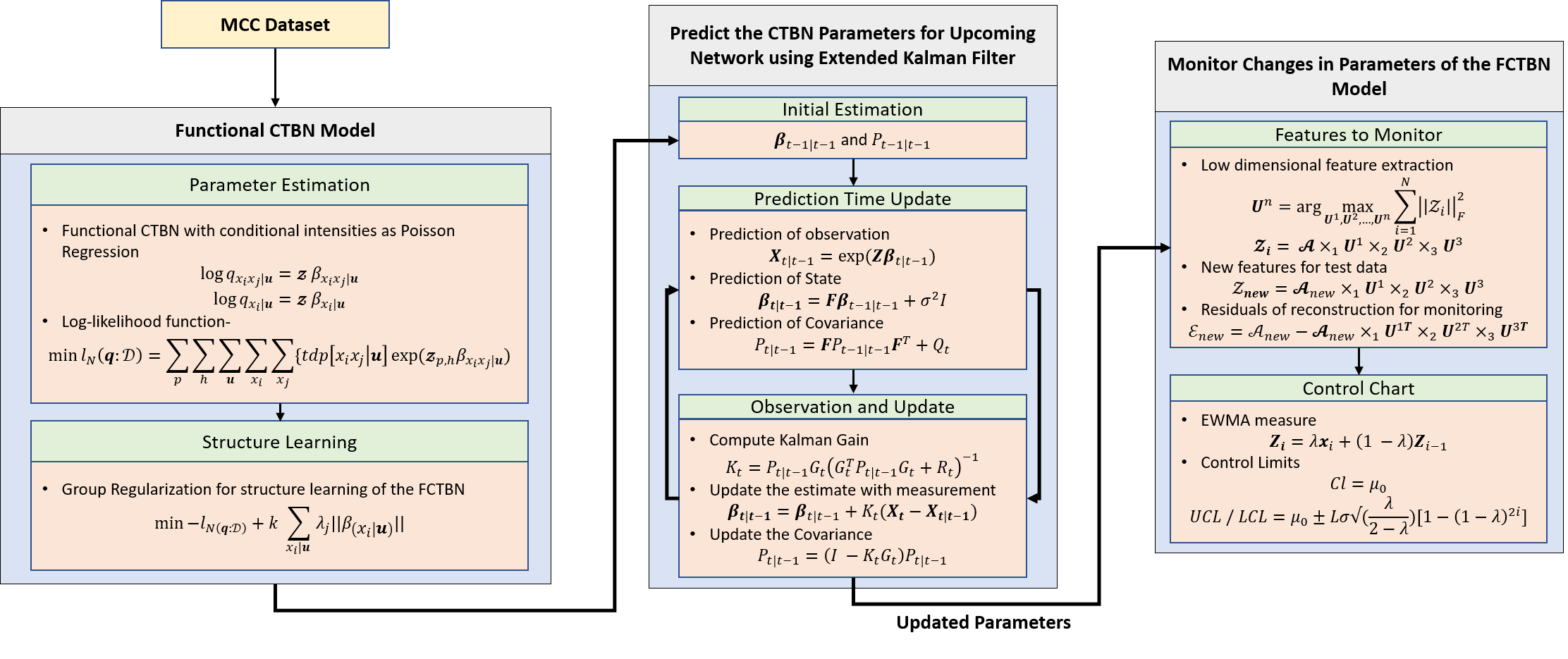}
{Overall schema of the proposed approach for dynamic prediction and monitoring of the emergence and progression of MCC.}\label{Figure:Overall_Planning}

The proposed methodology has the following contributions:
\begin{enumerate}
    \item We propose to formulate the conditional dependencies of FCTBN as a non-linear state-space model based on EKF to create a dynamic network  (D-FCTBN) that captures the dynamics of modifiable risk factors on the structure and parameters of the network.
    \item We propose a tensor control chart to monitor the evolution of the D-FCTBN network parameters over time proactively and signal when there is a significant change in the estimated parameters, which can result in an increased risk of developing new conditions.
    \item We validate the proposed methodology for dynamic prediction and monitoring of the emergence of multiple chronic conditions based on a combination of a simulated and real dataset from the Cameron County Hispanic Cohort Cohort (CCHC).
\end{enumerate}

The remainder of the paper is structured as follows. 
Section \ref{Section: Relevant_Background} presents the preliminaries and background for the CTBN and FCTBN.  Specifically, Section \ref{Subsection:Continuous_Time_Bayesian_Network} describes the details of the CTBN, and Section \ref{Subsubsection:Functional_CTBN} explains the functional CTBN and the regularized regression model for learning its structure and parameters. 
Section \ref{Section:Proposed_Approach} details the proposed approach for developing the Dynamic FCTBN (D-FCTBN) and the tensor control chart for monitoring the evolution of D-FCTBN. In particular, Section \ref{Subsection:Extended_Kalman_filter_Main} describes the details of the proposed EKF model for modeling the dynamics of edges of the D-FCTBN based on the changes in the modifiable risk factors and their interaction with existing conditions.
Also, Section \ref{Subsection:Monitoring_control_events} describes the building blocks of the proposed tensor control chart for monitoring the estimated parameters of the proposed D-FCTBN.
Section \ref{Section:Results_and_Discussion} presents the study population, the resulting model structure and parameters, and the tensor control chart to detect network changes. Finally, Section \ref{Section:Conclusion} provides the concluding remarks.

\section{Relevant Background}
\label{Section: Relevant_Background}
In this section, we review some of the major components of the proposed approach, including the CTBN for modeling MCC evolution as a finite-state continuous-time conditional Markov process over a factored state 
\cite{nodelman_continuous_2012, nodelman_expectation_2012, nodelman_learning_2012}, and functional CTBN (FCTBN) \cite{faruqui_functional_2020} for extending CTBN edges based on Poisson regression of some exogenous risk factors.

\subsection{Continuous Time Bayesian Network (CTBN)}
\label{Subsection:Continuous_Time_Bayesian_Network}

\subsubsection{CTBN Components}
\label{Subsubection:CTBN1}
Continuous time Bayesian networks (CTBNs) are Bayesian networks that models time explicitly by defining a graphical structure over continuous time Markov processes \cite{nodelman_continuous_2012}. 
Let $X = \{x_1,x_2,....,x_n\}$ denotes the state space of a set of random variables with 
discrete states $x_i=\{1,...,l\}$, such as MCC like Diabetes, Obesity, Hypertension, Heyperlipidemia, and Cognitive Impairment.
A CTBN consists of a set of conditional intensity matrices (CIM) under a given graph structure \cite{nodelman_continuous_2012, norris_markov_1998}. The components of a CTBN are - 

\begin{enumerate}
  \item An initial distribution $(P_x^0)$, which formulates the structure of the (conditional) relationship among the random variables and is specified as a Bayesian network, where each edge $x_i \to x_j$ on the network implies the impact of the parent condition $x_i$ on the child condition $x_j$.
  \item A state transition model $(Q_{X_i|\textbf{u}})$, which describes the transient behavior of each variable $x_i\in X$ given the state of parent variables $\textbf{u}$, and is specified based on CIMs -
\end{enumerate}
\[
\textbf{Q}_{X|\textbf{u}} = \begin{bmatrix} 
    -q_{x_1|\textbf{u}}    & q_{x_1x_2|\textbf{u}}    & \dots & q_{x_1x_n|\textbf{u}} \\
    q_{x_2x_1|\textbf{u}}  & -q_{x_2|\textbf{u}}      & \dots & q_{x_2x_n|\textbf{u}} \\
    \vdots      & \vdots        & \ddots & \vdots\\
    q_{x_nx_1|\textbf{u}}  & q_{x_nx_2|\textbf{u}}    & \dots & -q_{x_n|\textbf{u}}
    \end{bmatrix}
\]
\noindent where $q_{x_ix_j|\textbf{u}}$ represents the intensity of the transition from state $x_i$ to state $x_j$ given a parent set of node $\textbf{u}$, and $q_{x_i} = \sum_{j \neq i} q_{x_ix_j}$.  
Conditioning the transitions on parent conditions sparsifies the intensity matrix considerably, which is especially helpful for modeling large state spaces. When no parent variable is present, the CIM will be the same as the classic intensity matrix.

The probability density function ($f$) and the probability distribution function ($F$) for staying at the same state (say, $x_i$), which is exponentially distributed with parameter $q_{x_i}$, are calculated as-

\begin{align}
\label{Equation:Probability density}
    f(q_x,t) &= q_{x_i} exp(-q_{x_i}t),&{t \geq 0}\\
    F(q_x,t) &= 1 - exp(-q_{x_i}t),    &{t \geq 0}
\end{align}
\subsubsection{CTBN Parameter Estimation}
\label{Subsubsection:Parameter_Estimation1}
Given a dataset $\mathcal{D} = \{\tau_{h=1}, \tau_{h=2},....,\tau_{h=H}\}$ of $H$ observed transitions, where $\tau_h$ represents the time at which the $h^{th}$ transition has occurred, and $\mathcal{G}$ is a Bayesian network defining the structure of the (conditional) relationship among variables, we can use maximum likelihood estimation (MLE) (equation \eqref{Equation:Likelihood_Function}) to estimate parameters of the as defined in Nodelman et al \cite{nodelman_continuous_2012, nodelman_learning_2012}-

\begin{align}
\label{Equation:Likelihood_Function}
    \begin{split}
    L_x(q_{x|\textbf{u}}:\mathcal{D}) &= \prod_{\textbf{u}} \prod_x q_{x|\textbf{u}}^{M{[x|\textbf{u}]}} exp(-q_{x|\textbf{u}}T{[x|\textbf{u}]})
    \end{split}
\end{align}

\noindent where, $T[x|\textbf{u}]$ is the total time $X$ spends in the same state $x$, and $M{[x|\textbf{u}]}$ the total number of time $X$ transits out of state $x$ given, $x = x'$ .The log-likelihood function can be then written as- 
\begin{align}
\label{Equation:Log-Likelihood_Function}
    \begin{split}
    l_x(q_{x|\textbf{u}}:\mathcal{D}) &= \sum_{\textbf{u}} \sum_x M[x|\textbf{u}]\hspace{1pt} ln(q_{x|\textbf{u}}) - q_{x|\textbf{u}}\hspace{1pt} T[x|\textbf{u}]
    \end{split}
\end{align}
Maximizing equation \eqref{Equation:Log-Likelihood_Function}, provides the maximum likelihood estimate of the paramters of the FCTBN.

\subsection{Functional CTBN (FCTBN)}
\label{Subsubsection:Functional_CTBN}

\subsubsection{FCTBN with Conditional Intensities as Poisson Regression}
\label{Subsubection:FCTN_PoissonRegression}
In reality, the progression of state variables, such as chronic conditions, not only depends on the state of their parents, such as preexisting chronic conditions but some exogenous variables, such as patient level risk factors like age, gender, etc. 
Using Poisson regression to represent the impact of exogenous variables on the conditional dependencies, the rate of transition between any pair of MCC states can be derived as \cite{faruqui_functional_2020}-
\begin{subequations}
    \begin{align}
            \label{Equation:CTBN_Poisson_1}
        \log {q}_{x_i,x_j|\textbf{u}} &= \beta_{0_{x_i,x_j|\textbf{u}}} + \beta_{1_{x_i,x_j|\textbf{u}}} + ... ... + z_m \beta_{m_{x_i,x_j|\textbf{u}}} \\
                        &= \boldsymbol{z} \boldsymbol{\beta}_{\textbf{x}_i,\textbf{x}_j|\textbf{u}}
    \end{align}
\end{subequations}

\noindent where, $\textbf{z}=\{z_1,z_2,...,z_m\}$ is the set of exogenous variables (e.g. patient-level risk factors such as age, gender, race, education, marital status, etc.), and $\beta_{k_{x_i|\textbf{u}}}=\sum_{j\neq i}\beta_{k_{x_ix_j|\textbf{u}}}, k=0,\ldots,m$ is the set of coefficients (parameters) associated with the exogenous variables. 
Also, the rate of staying in the same state is modeled as-
\begin{subequations}
    \begin{align}
        \label{Equation:CTBN_Poisson_2}
        \log q_{x_i|\textbf{u}} &= \beta_{0_{x_i|\textbf{u}}} + z_1 \beta_{1_{x_i|\textbf{u}}} + ... ... + z_m \beta_{m,{x_i|\textbf{u}}} \\
                        &= \boldsymbol{z} \boldsymbol{\beta}_{\textbf{x}_i|\textbf{u}}
    \end{align}
\end{subequations}

When the state space of the random variables is binary, as in our case study on MCC transitions, where MCC states include having/not having each of the conditions, the conditional intensities in $\textbf{Q}_{x_i |\textbf{u}_i}$, can be estimated just using Equation  \ref{Equation:CTBN_Poisson_2} because for Markov processes with binary states $ q_{x_i |\textbf{u}} = - \sum_{j\neq i} q_{(x_i x_j |\textbf{u})}$. 
This feature considerably simplifies the estimation of the functional CTBN conditional intensity matrix based on Poisson regression.
\subsubsection{Parameter Estimation}
\label{Subsubsection:Parameter_Estimation}
Having the dataset $D=\left\{\tau_{\left(p=1,h=1\right)},\ldots,\tau_{\left(P,H\right)}\right\}$ of MCC trajectories, where $\tau_{\left(p,h\right)}$ represents the time at which the $h^{th}$ (MCC) transition of  the $p^{th}$ patient has occurred, we use maximum likelihood estimation to estimate parameters of the proposed FCTBN. 
The likelihood of $D$ can be decomposed as the product of the likelihood for individual transitions. Let $d=\langle \textbf{z},\textbf{u},x_i|\textbf{u},t_d,x_j|\textbf{u} \rangle$ be the transition of patient $p$ with risk factors $\textit{z}$ and existing conditions $\textbf{u}$, who made the transition to state $x_{j|\textbf{u}}$ after spending the amount of time $t_d=\tau_{\left(p,h\right)}-\tau_{\left(p,h-1\right)}$ in state $x_{i|\textbf{u}}$. 
By multiplying the likelihoods of all conditional transitions during the entire trajectory for all patients $p=1,\ldots,P$, and taking the log, we obtain the overall log-likelihood function as- 
\begin{align}
\label{Equation:FCTBNLikelihood_Function}
    \begin{split}
    l_N\left(\textbf{q}:\mathcal{D}\right)\
    &=\sum_{p}\sum_{h}\sum_{\textbf{u}}\sum_{x_j}\sum_{x_i} \\
    & \left\{{t_d}_p\left[x_ix_j|\textbf{u}\right]\exp{\left(\boldsymbol{z}_{p,h}\boldsymbol{\beta}_{x_ix_j|\textbf{u}}\right)}\right\}
    \end{split}
\end{align}
which is a convex function and can be maximized efficiently using a convex optimization algorithm such as Newton Raphson to estimate parameters $\textit{\beta}_{x_i|\textbf{u}}$. 

\subsubsection{Group regularization for structure learning of the FCTBN}
\label{Subsubsection:Group regularization}
The parameter estimation approach presented above requires the parent set of each condition to be known, which is equivalent to knowing the Bayesian network structure. Given that FCTBN has a special structure based on a conditional intensity matrix that allows for cycles, group regularization can be used to penalize groups of parameters pertaining to each specific conditional transition (each edge) \cite{faruqui_functional_2020} as-
\begin{equation}
    \label{Equation:CTBN_Minimization_1}
    \centering
    \min -l_N(\textbf{q}:\mathcal{D}) + k \sum_{x_{i}|\textbf{u}}\lambda_j\|\boldsymbol{\beta}_{x_{i}|\textbf{u}}\|
\end{equation}

\noindent where, $\|\boldsymbol{\beta}_{x_{i}|\textbf{u}}\| = \sqrt{\sum_{\textbf{u}} \sum_{x_{i}} (\boldsymbol{\beta}_{x_{i|\textbf{u}}}.\boldsymbol{\beta}^T_{x_{i|\textbf{u}} )}}$ is the $L_1$-norm of the group of parameters associated with each conditional transition. $k$ is the groups size which is based on the number of coefficients in the Poisson regression for each conditional intensity. $\lambda_j = \lambda \|\Tilde{\boldsymbol{\beta}_j}\|^{-1}$ is the tuning parameters (of the adaptive group regularization) that control the amount of shrinkage, where $\lambda$ is inversely weighted based on the unpenalized estimated value of the regression coefficients $\Tilde{\boldsymbol{\beta}_j}$  \cite{wang_note_2008}.\\

\section{Proposed Approach}
\label{Section:Proposed_Approach}
In this section, we first propose an extended Kalman filter to capture the effects of the dynamics of modifiable risk factors on the parameters, edges, and structure of the FCTBN (D-FCTBN). Next, we develop a tensor control chart to monitor the evolution of the dynamic FCTBN (D-FCTBN).

\subsection{An extended Kalman filter for dynamic prediction of FCTBN parameters}
\label{Subsection:Extended_Kalman_filter_Main}
The conditional dependencies (edges) of FCTBN  provide the rate of transitioning from one state to another given the parents' state and exogenous variables, i.e., the rate of a new chronic condition such as obesity emergence during the next $t$ years given the preexisting conditions such as diabetes and patient's level risk factors such as gender, age, etc. 
However, in reality, conditional dependencies dynamically change based on a person's modifiable behavioral factors, i.e., diet, exercise, and interaction with non-modifiable risk factors and existing conditions. To capture the dynamics of the changes in the conditional intensities (risk) of MCC, we propose to transform the parameters (coefficients) of the regression functions, which represent the edges of the MCC (FCTBN) network, into an extended Kalman filter (EKF) \cite{gahrooei_change_2018}. 

EKF consists of an observation equation and a state transition equation. The observation equation describes the most recent observation of state variables using system dynamics, namely Poisson regression coefficients associated with the emergence and progression of MCC. The transition equation predicts how the state variables evolve to the next period, namely how the coefficients associated with MCC will progress/emerge in the next period (See Fig. \ref{Figure:EKF_Illustration}).

\noindent \textit{\textbf{Observation equation:}} Each edge/connecting in the MCC FCTBN network represents, the rate of occurrence of a chronic condition such as diabetes, based on a Poisson regression function with parameters $\left[\textit{\beta}_{x_ix_j|\textit{u}}\right]_t$. We consider $\left[\textit{\beta}_{x_ix_j|\textit{u}}\right]_t$ as the state variables of the dynamical system which describe the (noisy) sequences of MCC observations. This results in the observation equation given by-
\begin{align}
    \label{Equation:Observation_0}
    \left[q_{x_i|\textbf{u}}\right]_t=exp\left(\boldsymbol{z}_\textit{t}\left[\boldsymbol{\beta}_{x_i|\textbf{u}}\right]_t\right)\ 
\end{align}
The observation equation \eqref{Equation:Observation_0} is non-linear and thus we will employ the extended Kalman filter (EKF) \cite{fahrmeir_kalman_1991} instead of general Kalman filters (KF) \cite{kalman_new_1960}. EKFs similar to KF follows a recursive procedure where it performs predictions based on a given observation and updates the estimates iteratively \cite{fahrmeir_kalman_1991}. 

\noindent \textit{\textbf{State transition equation:}} As a patient changes her lifestyle behaviors, the state variables of the proposed dynamical model evolve in time to best predict the MCC emergence and progression. This results in a state transition equation given by-
\begin{align}
    \label{Equation:Statetransition_0}
    \left[\boldsymbol{\beta}_{x_i|\textbf{u}}\right]_t=\boldsymbol{F}\left[\boldsymbol{\beta}_{x_i|\textbf{u}}\right]_{t-1}+\textit{\varepsilon}_t
\end{align}
\noindent where $\boldsymbol{F}$ is the state transition matrix, and $\varepsilon_t$ is the white noise assumed to follow a Gaussian distribution with mean zero and covariance $\sigma^2\mathbf{I}$. The transition matrix $\boldsymbol{F}$ can be approximated from stream of data $\boldsymbol{X}_{i|t}$ utilizing the FCTBN model evaluated at different point in time or by utilizing some system identification techniques \cite{ljung_system_2017}.   
\Figure[t!](topskip=0pt, botskip=0pt, midskip=0pt)[scale = 0.2]{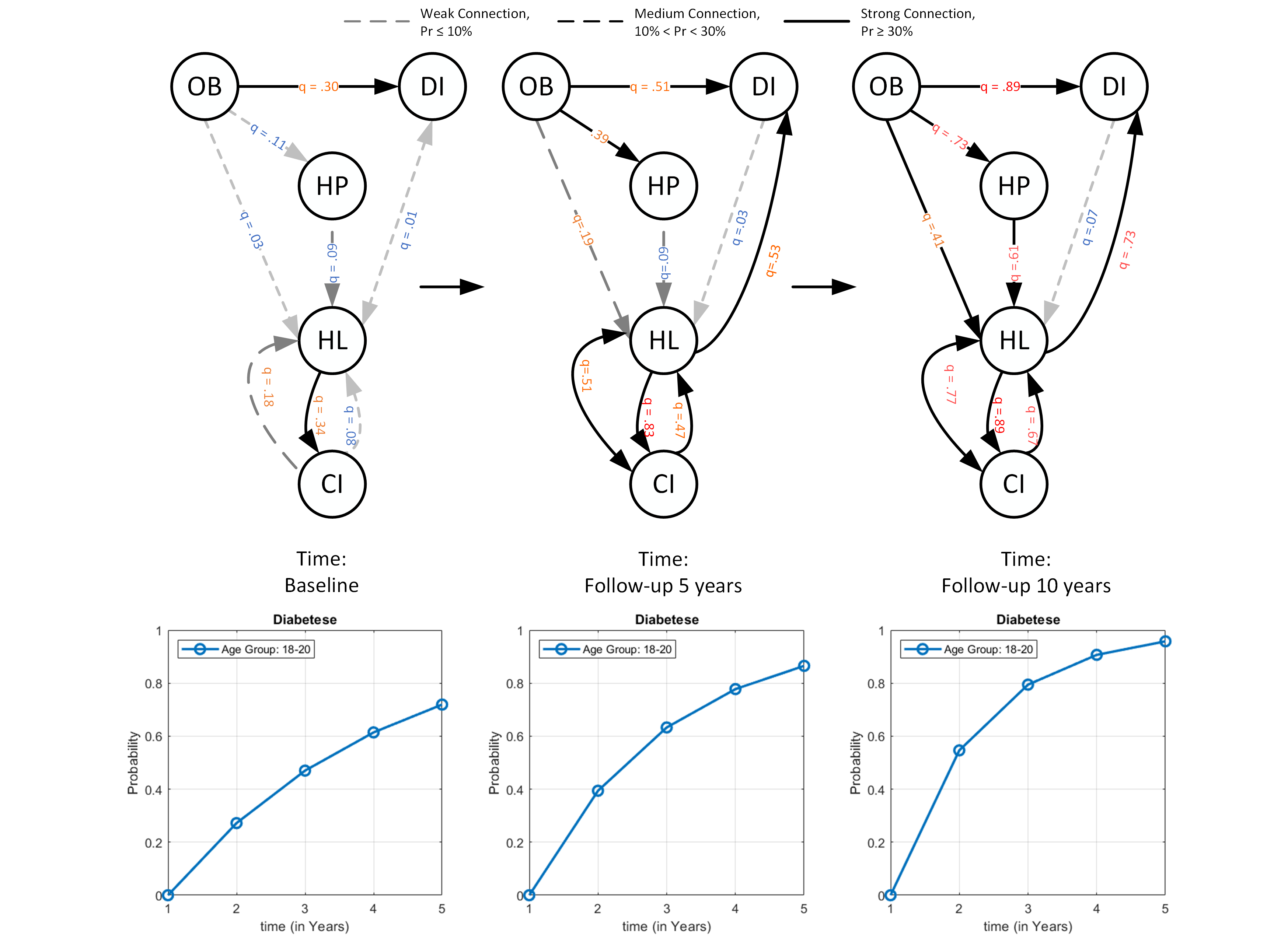}
{Illustration of the impact of behavioral risk factors dynamics on the conditional intensities/dependencies and risk trajectory of developing new MCC conditions, i.e. Diabetes, at three time points, including baseline, 5-year follow up, and 10-year follow up, using extended Kalman filter; The nodes with thick outlines represent the existing or developed conditions over time. (The nodes, OB: Obesity, HP: High Blood Pressure, DI: Diabetes, HL: Hyperlipidemia, and CI: Cognitive Impairment).\label{Figure:EKF_Illustration}}


\noindent\textit{\textbf{Dynamic Prediction via EKF:}} EKF takes the most recent estimate of the state variables with information of changes in the (modifiable) behavioral risk factors up to time $t$ and uses the system dynamics to predict the future state of the variables and prediction of the MCC as \cite{fahrmeir_kalman_1991}-
 
\begin{align}
    \label{Equation:Prediction_1}
    [\boldsymbol{\beta}_{x_i|\textbf{u}}]_{t|t-1} &= \boldsymbol{F}[\boldsymbol{\beta}_{x_i|\textbf{u}}]_{t-1|t-1} + \sigma^2 I\\
    \label{Equation:Prediction_2}
    P_{t|t-1} &= \boldsymbol{F}[\boldsymbol{\beta}_{x_i|\textbf{u}}]_{t-1|t-1}\boldsymbol{F}^T + Q_t
\end{align}

where $[\boldsymbol{\beta}_{x_i|\textbf{u}}]_{t|t-1}$ and $P_{t|t-1}$ are the extended Kalman prediction of the matrix of estimated coefficients and their covariance 
respectively given a set of observations $\boldsymbol{X}_{i,t}$. The observation equation is linearized using the Taylor Expansion to achieve a sub-optimal estimate of the state value. 
\\
\\
\noindent\textit{\textbf{Dynamic estimation via EKF:}} When new observations of MCC are obtained, the error between the observation and the EKF predictions is used to update the posterior mean of the state variable as-
\begin{align}
    \label{Equation:Update_1}
    [\boldsymbol{\beta}_{x_i|\textbf{u}}]_{t|t-1} &= [\boldsymbol{\beta}_{x_i|\textbf{u}}]_{t|t-1} + K_t (\boldsymbol{X_{x_i|\textbf{u},t}} - \boldsymbol{X}_{x_i|\textbf{u},t|t-1})
\end{align}

\noindent where $K_t$ is the Kalman gain and calculated using the following equations-
\begin{align}
    \label{Equation:Update_2}
    P_{t|t-1} &= (I - K_t G_t) P_{t|t-1}\\
    \label{Equation:Update_3}
    K_t &= P_{t|t-1} G_t (G_t^T P_{t|t-1} G_t + R_t ) ^{-1}
\end{align}

\noindent where, $G_t$ denotes the Jacobian of $g$ evaluated at $[\boldsymbol{\beta}_{x_i|\textbf{u}}]_{t|t-1}$ i.e. $G_t = \frac{\partial{g}}{\partial{[\boldsymbol{\beta}_{x_i|\textbf{u}}]_t}}|_{[{\boldsymbol{\beta
}_{x_i|\textbf{u}}]}_{t|t-1}} = \boldsymbol{Z}_t^T diag(exp(\boldsymbol{Z}_t[\boldsymbol{\beta}_{x_i|\textbf{u}}]_{t|t-1})) $, $R_t$ represents the variance of observations and is estimated based on the underlying network distribution and the observation prediction. The estimated parameter $[\boldsymbol{\beta}_{x_i|\textbf{u}}]_{t|t}$ provides a sub-optimal estimate of the network parameters at time $t$ \cite{fahrmeir_kalman_1991}.



\subsection{Monitoring of events}
\label{Subsection:Monitoring_control_events}
In this section, we propose a monitoring scheme to determine meaningful changes in the exogenous variables (modifiable risk factors) that can have an impact on the risk of developing new chronic conditions. For this purpose, we propose a statistical control chart that automatically signals when there is a meaningful change in the predicted value of the coefficients associated with the patient level (modifiable) risk factors, namely $[\boldsymbol{\beta}_{x_i|\textbf{u}}]_{t|t-1}$, which is dynamically updated by the D-FCTBN. The idea behind monitoring the $[\boldsymbol{\beta}_{x_i|\textbf{u}}]_{t|t-1}$ is that the predicted value of the risk factors coefficients are directly related to the network edges (conditional intensities) and the risk of developing new MCC conditions.

Given the dynamic prediction of D-FCTBN (MCC network) parameters for a time point ${t|t-1}$, the coefficients form a 3-dimensional tensor including the parents, children, and risk factors dimensions (modes). To effectively monitor the tensor of predicted coefficients for any potential changes, we propose to use multilinear principal component analysis (MPCA) to extract the most salient features of the data for building the control chart. It is worth mentioning that, there are  other alternative methods that can also be used for extracting major features of the coefficients tensor, including Unfolded  PCA \cite{liu_extraction_2007, bharati_image_2004},  Multilinear Principal Component Analysis (MPCA) \cite{lu_mpca_2008}, Uncorrelated MPCA \cite{lu_uncorrelated_2009}, Robust MPCA \cite{dehaan_adaptive_2007, inoue_robust_2009} , and Non-negative MPCA  \cite{zass_nonnegative_2006}. Interested readers may also refer to Kruger et al \cite{kruger2008developments}, Lu et al \cite{lu_mpca_2008} and Paynabar et al \cite{paynabar2013monitoring} for comprehensive reviews of tensor feature extraction methods.

\subsubsection{Multilinear Principal Component Analysis}
\label{subsubsection:Multilinear_Principal_Component_Analysis}
Lu et al. \cite{lu_mpca_2008} introduce the MPCA framework for tensor feature extraction. They decompose the original tensor into a series of multiple projection sub-problems and solves them iteratively.  

\Figure[h!](topskip=0pt, botskip=0pt, midskip=0pt)[width=.45\textwidth]{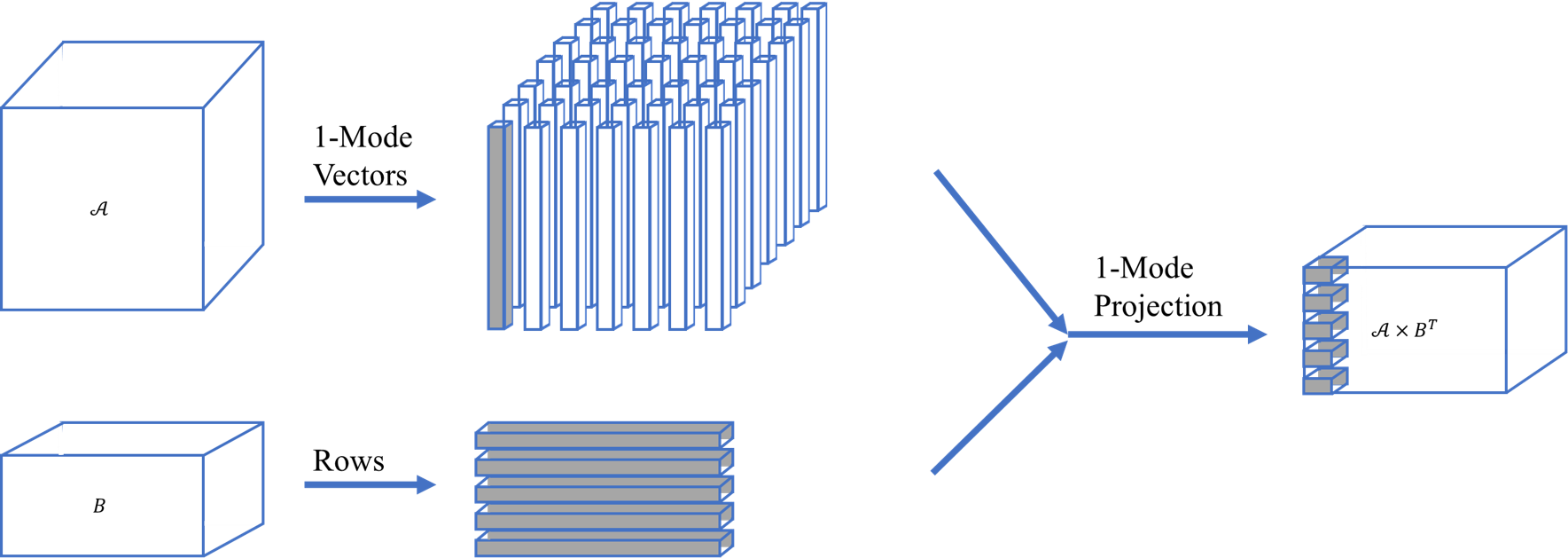}
{Visual illustration of multilinear projection; projection in the 1-mode vector space.\label{Figure:MPCA_Illustrated2}}

For an in-control training set of $N$-th order tensor denoted as $\mathcal{A} \in \mathbb{R}^{I_1 \times I_2, \times I_3, ... ..., \times I_N}$, the MPCA projection can be denoted by $y \in \mathbb{R}^{I_1 \times I_2 \times I_3}$, where $n = 1, 2, ... ..., N$, $I_1, I_2$ and $I_3$ are the dimensions of the coefficient tensor and $N$ is the number of updated coefficients attained from EKF. 
Lu et al. \cite{lu_mpca_2008} find the set of orthogonal transformation matrices, 
$\textbf{U}^n \in \mathbb{R}^{I_n \times P_n}, n = 1, 2, 3, ...., N$, 
where the dimensionality $P_n$ ($P_n \leq I_n$) for each mode is predetermined or known for the application of interest. 
They have also developed methods for adaptive determination of $P_n$ in case it's not pre-determined. The transformation is performed such that it captures the most variations of the original tensor. To keep the original estimated values of the coefficients, in this work we utilized the non-centered version of the MPCA. Therefore, the low dimensional features after applying MPCA will be,

\begin{equation}
    \textbf{U}^n = \arg\max_{{{\bf U}}^{(1)},{{\bf U}}^{(2)},\ldots,{{\bf U}}^{(N)}} \sum_{i=1}^N || \mathcal{Z}_i ||^2_F
\end{equation}

\noindent where, $n = 1, 2, 3$, $\mathcal{Z}_i = \mathcal{A} \times_1\textbf{U}^{(1)} \times_2\textbf{U}^{(2)} \times_3\textbf{U}^{(3)}$ and $\mathcal{A}$ is non-centered tensor data. In case of centered data, $\mathcal{A}$ can be replaced with $\widetilde{\mathcal{A}}$, where $\widetilde{\mathcal{A}} = {\mathcal{A}} - \overline{\mathcal{A}}$. For a new feature tensor, $\mathcal{A}_{new} \in \mathbb{R}^{I_1 \times I_2, \times I_3}$, the features are calculated as,  
\begin{equation}
    \centering
    \label{Equation:MPCA_New_Feature}
    \mathcal{Z}_{new} = \mathcal{A}_{new} \times_1 \textbf{U}^{(1)} \times_2 \textbf{U}^{(2)} \times_3 \textbf{U}^{(3)}
\end{equation}
and residuals of reconstruction can be calculated as 
\begin{equation}
    \centering
    \label{Equation:MPCA_New_Error_Feature}
    \mathcal{E}_{new} = \mathcal{A}_{new} - \mathcal{Z}_{new} \times_1 \textbf{U}^{(1)T} \times_2 \textbf{U}^{(2)T} \times_3 \textbf{U}^{(3)T}
\end{equation}

The errors at every time step can also be vectorized by calculating the \textit{norm} of all the data, i.e. $\mathcal{E}_{new\_vector} = ||\mathcal{E}_{new}||_2$.

\subsubsection{Monitoring Scheme}
\label{subsubsection: Monitoring_Scheme}
Here, we propose a tensor control chart to monitor the changes in the D-FCTBN network edges caused by changes in the patient modifiable risk factors, namely lifestyle behavioral changes. Given the estimate of FCTBN parameters $\sigma^2 I$ and $[\boldsymbol{\beta}_{x_i|\textbf{u}}]_{t|t-1}$ based on Section \ref{Subsubsection:Functional_CTBN}, for any new observation of patients (modifiable and non modifiable) risk factors and MCC conditions, the tensor of new network parameters (risk factors' coefficients) are predicted using the EKF detailed in Section \ref{Subsection:Extended_Kalman_filter_Main}, and the relevant features are extracted using MPCA discussed in Section \ref{subsubsection:Multilinear_Principal_Component_Analysis}.

When there is no significant change in the patients' lifestyle behaviors, the reconstruction error in Equation \ref{Equation:MPCA_New_Error_Feature} will be small, as patients' historical/past behavior can accurately estimate the F-FCTBN parameters. However, when there is a significant change in the patients' lifestyle behaviors, the distribution of reconstruction error will change, and the observed value will supposedly increase. Therefore, for new predictions of the D-FCTBN parameters,
$[\boldsymbol{\beta}_{x_i|\textbf{u}}]_{t+1|t} \approx \mathcal{A}_{new} \in \mathbb{R}^{I_1 \times I_2, \times I_3}$, the reconstruction error can be used to identify potential high-impact changed in patients modifiable risk factors. The proposed monitoring scheme is based on a Multivariate Exponential Weighted Moving Average (MEWMA) \cite{lowry_multivariate_1992} control chart of the vectorized reconstruction error:
\begin{equation}
    \centering
    \label{Equattion: MEWMA_Smoothing}
    \boldsymbol{Z}_i = \lambda \textbf{x}_i + (1 - \lambda) \boldsymbol{Z}_{i-1}
\end{equation}
\noindent where $0 \leq \lambda \leq 1$ and $\boldsymbol{Z}_0 = 0$. In case of MEWMA the quantity plotted on the control chart is -
\begin{equation}
    \centering
    \label{Equation:T2}
    T_i^2 = \boldsymbol{Z}_i^T S_I^{-1} \boldsymbol{Z}_i
\end{equation}
\noindent where the covariance matrix is, 
\begin{equation}
    \centering
    \label{Equation:Covariance}
    S_I^{-1} = \frac{\lambda}{2 - \lambda} [1 - (1 - \lambda)^{2i}] S
\end{equation}

\noindent which is equivalent to the variance of the univariate EWMA, and $S$ is the sample covariance matrix calculated of the features estimated by $N$ in-control samples \cite{montgomery_introduction_2020}. The control limits of control chart can be calculated as follows (for the univariate case)-

\begin{align}
    \begin{split}
        \label{Equation: Limits_of_EWMA}
            UCL &= \mu_0 + L \sigma \sqrt{\frac{\lambda}{(2 - \lambda)} [1 - (1 - \lambda)^{2i}]} \\
            CL  &= \mu_0 \\
            LCL &= \mu_0 - L \sigma \sqrt{\frac{\lambda}{(2 - \lambda)} [1 - (1 - \lambda)^{2i}]}        
    \end{split}
\end{align}

\noindent Where, $L$ is the width of the control limits, $\sigma^2$ is the variance of the data, and $i$ represents the observation number of the EWMA statistics.
The selection of the features to be used is determined based on the percentage of the total variance explained by the extracted features using MPCA.

\section{Results and Discussion}
\label{Section:Results_and_Discussion}
Long-lasting diseases, otherwise known as chronic conditions, can be considered a degradation process that progresses over time and contributes to the development of other new chronic conditions. The presence of two or more chronic medical conditions in an individual is commonly defined as multimorbidity, or multiple chronic conditions (MCC) \cite{faruqui_mining_2018, pugh_complex_2014}. Here, we use the proposed dynamic FCTBN (D-FCTBN) to find the impact of patient level risk factors, specifically lifestyle behaviors, on the conditional dependencies of MCC over time. In addition, we use the proposed tensor control chart to monitor the risk of new MCC emergence based on the dynamics of patients' lifestyle behaviors. 

\subsection{Study Population and Demographics}
\label{Subsection:Study_Population}
\Figure[h!](topskip=0pt, botskip=0pt, midskip=0pt)[width=.45\textwidth]{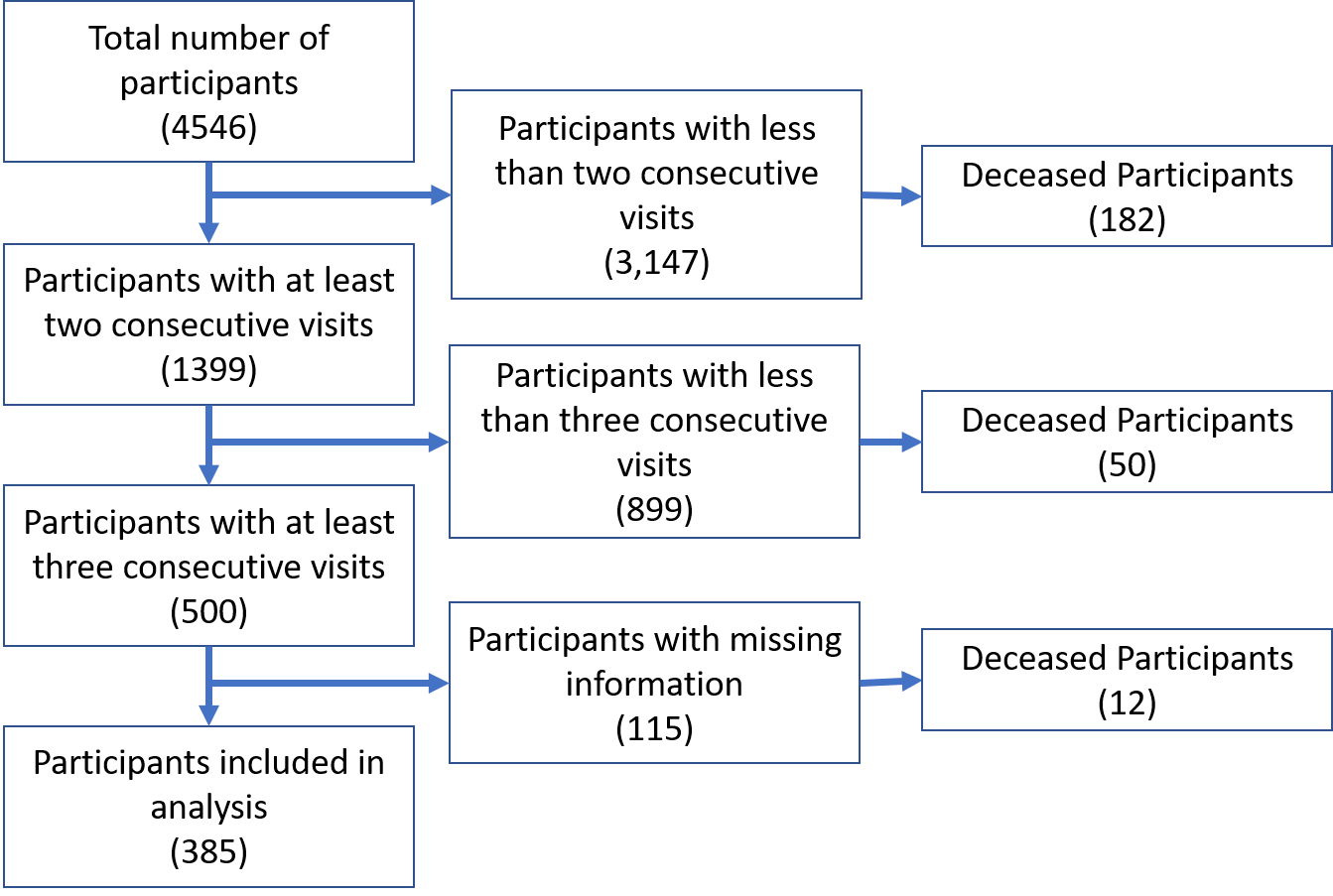}
{Flow diagram of sample selection and the final number of patients included in the analysis.\label{Figure:Study_Population_Selection}}

Our case study is based on the Cameron County Hispanic Cohort (CCHC) dataset for comorbidity analyses. The CCHC is a cohort study comprised of mainly Mexican Americans (98\% of cohort) randomly recruited from a population with severe health disparities on the Texas-Mexico border and started in 2004. 
The CCHC is employing a rolling recruitment strategy and currently numbers 4,546 adults. Inclusion criteria: (1) participating in the study between 2004 and 2020, (2) having at least three 5-year follow-up visits during that period. 385 patients met these criteria, which include the dataset of our study (see Figure \ref{Figure:Study_Population_Selection}). The survey includes participants’ socio-demographic factors (age, gender, marital status, education, etc.) and lifestyle behavioral factors (diet, exercise, tobacco use, alcohol use, etc.). 

\subsection{Diagnosed Health Condition and Patient Associated Risk Factors}
\label{Subsection:Study_Population_2}
For this study, we considered some of the most common MCCs present in the Hispanic community, including diabetes, obesity, hypertension, hyperlipidemia, and mild cognitive impairment. The positive criteria (considering the condition to be active) for the conditions selected as below-
\begin{itemize}
    \item \textbf{Diabetes}: Fasting Glucose >= 126 mg/dL, HbA1c >= 6.5\%, or take diabetes medication \cite{association_diagnosis_2010}.
    \item \textbf{Obesity}: Body mass index (BMI, kg/m2)>= 30 \cite{fruhbeck_abcd_2019}.
    \item \textbf{Hypertension}: Systolic blood pressure (BP) >= 130 mmHg, Diastolic BP >= 80 mmHg, or take anti-hypertensive medication \cite{pk_2018}.
    \item \textbf{Hyperlipidemia}: Total cholesterol > 200 mg/dL, triglycerides >= 150 mg/dLl, HDLC < 40 mg/dL (for male)/ HDLC < 50 mg/dL (for female), LDLC >= 130 mg/dL, or take medication for hyperlipidemia \cite{grundy_scott_m_2018_2019}.
    \item \textbf{Mild Cognitive Impairment}: Mini-Mental State Score < 23 (out of 30) \cite{wu_association_2018}.
\end{itemize}

\Figure[h!](topskip=0pt, botskip=0pt, midskip=0pt)[width=1\textwidth]{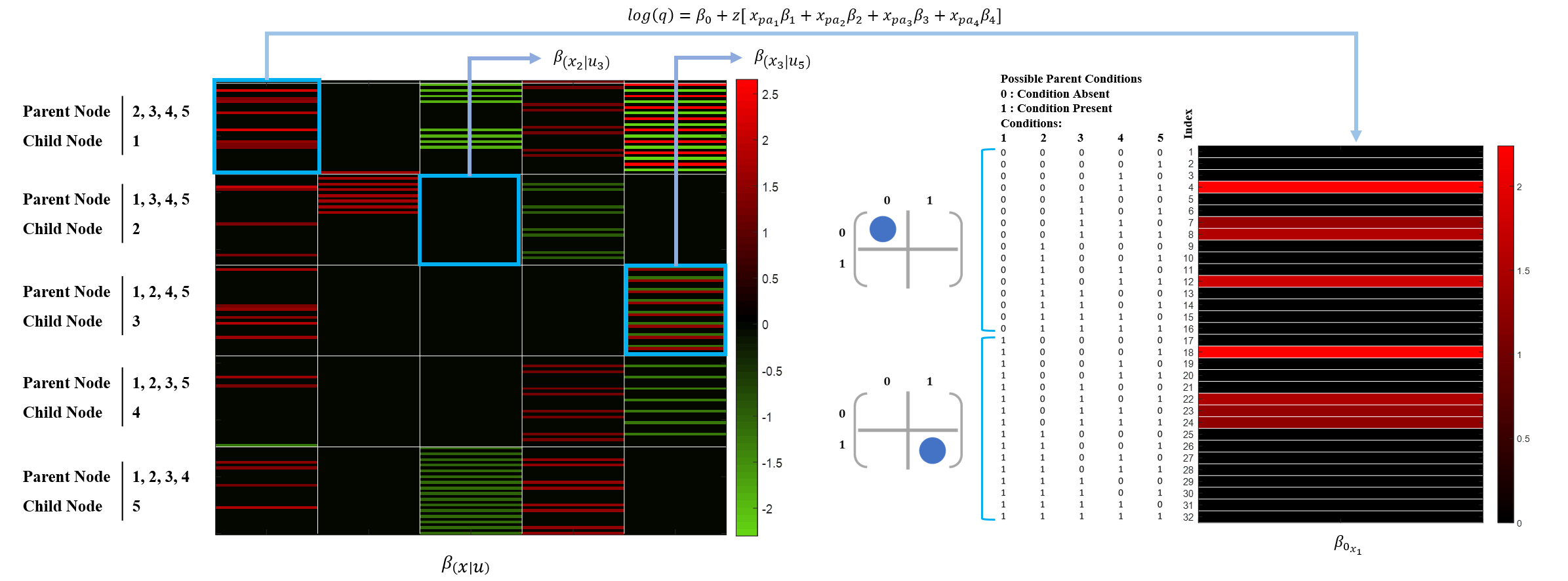}
{The estimated parameters of FCTBN based on the optimal value of tuning parameters. The matrix contains all the possible combinations of parent and child interaction. For example, the first row set (first 32 rows) of the matrix represents the parameters learned child node 1 while considering the parents' node are 2, 3, 4, and 5. The right side of the Figure shows all the possible condition possible (1 for the presence of a condition and 0 for no presence of no condition).\label{Figure:Transition_Explain}}

For the risk factors, the dataset includes the participant's non-modifiable risk factors based on socio-demographic information (age, gender, and education history) and modifiable risk factors based on lifestyle behavioral factors (diet, exercise, tobacco use, alcohol use). Diet and exercise are categorized according to the U.S. Healthy Eating Guideline, and U.S Physical Activity Guideline \cite{wu_fruit_2019}.

\subsection{FCTBN Structure and Parameter Learning}
\label{Subsection:Learned_Bayesian_Structure}
To identify the optimal value of the tuning parameter ($\lambda$) of the adaptive group regularization method for FCTBN structure and parameter learning, we use cross-validation error based on several $\lambda$ values. We attain the structure of FCTBN and the parameters using the optimal value of $\lambda = 10^2$. Figure \ref{Figure:Transition_Explain} provides the heatmap of the estimated parameters for the based FCTBN model. These learned parameters will be used as the initial parameters of EKF for estimating the dynamic FTCBN and monitoring possible changes in the risk of acquiring a new MCC condition.

\subsection{D-FCTBN Dynamic Estimation and Prediction Using EKF}
\label{Subsection:EKF_Update}
The estimated parameters of FCTBN provide the baseline/initial values of the D-FCTBN. As new (dynamic) observations of patient's (modifiable) risk factors and MCC status are made available, we use EKF to capture the dynamics of a patient's modifiable risk factors and MCC update, as detailed in Section \ref{Subsection:Extended_Kalman_filter_Main}. Figure \ref{Figure:Change_Coefficients}, visualizes the estimated parameter of the D-FCTNB for time, $t+1$ given the parameter information at time, $t$ and base parameter, $\beta_{t+1|t}$ using the proposed EKF module for 5 patients over 11 consecutive year. The proposed model provides a near-optimal approach to estimate and update the D-FCTBN parameters. The illustrated changes in D-FCTBN network parameters demonstrate the dynamics of patient's lifestyle behaviors and their impact on the MCC. 

\Figure[t!](topskip=0pt, botskip=0pt, midskip=0pt)[width=.9\textwidth]{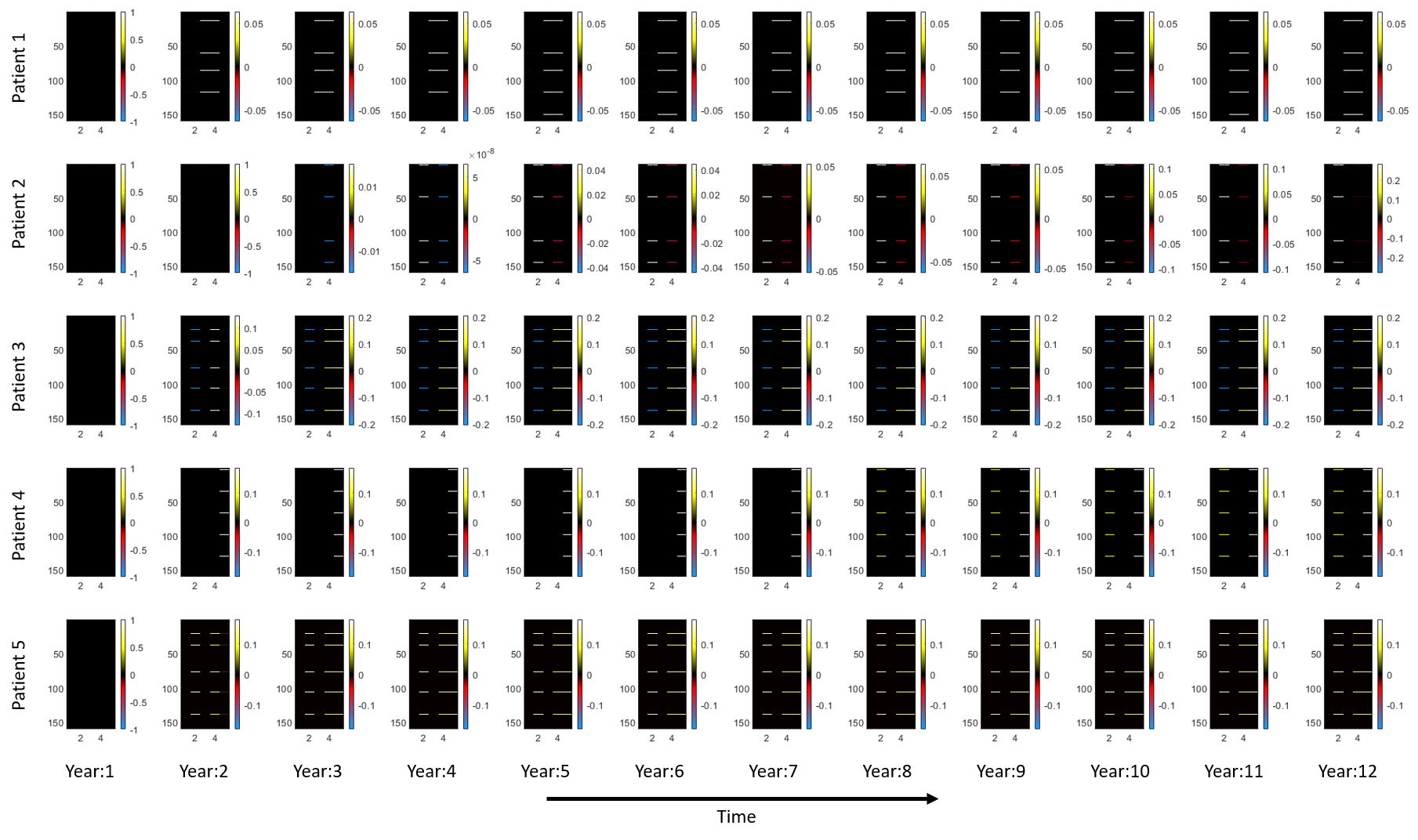}
{The EKF estimation of the parameters of the propsoed D-FCTBN for 5 patients patients over 11 consecutive periods. The illustration shows the change in parameters with respect to base year ($t=0$).\label{Figure:Change_Coefficients}}

\subsubsection{Stability Analysis of EKF for Estimating Parameters of FCTBN}
\label{Subsubsection:Stability_EKF}
In this section, we will discuss the stability of the EKF model derived in Section \ref{Subsection:Extended_Kalman_filter_Main}. For the measurement error, $\textbf{e} = \textbf{X}_t - \textbf{X}_{t|t-1}$ Konrad et al. \cite{reif1999stochastic} showed that the estimation error remains bounded if the following conditions hold-

\begin{enumerate}
    \item $||\textbf{F}(\textbf{X}_t) || \leq \alpha $, \\
          $||G_t(\textbf{X}_t) ||  \leq \beta $, \\
          where $\alpha, \beta \geq 0$ positive real number for each, $t$.
    \item \textbf{F} is non-singular for every $t$.
    \item The estimation error, $\textbf{e}$ is exponentially bounded in mean square error. This also bounds the probability to one. This is only true when the estimates satisfy the condition $||\textbf{e}|| \leq \epsilon$ and the covariance matrices of the noise terms are bounded via, $\sigma^2 I \leq \delta I$ and $QI \leq \delta I$, where $\delta , \epsilon > 0$. 
\end{enumerate}

\noindent EKF model generally needs additional steps to correct the estimation of the future state. These additional steps are necessary to make sure estimated parameters do not diverge over time.
We consider mean squared error (MSE) for stability analysis of EKF.
Figure \ref{Figure:MSE_Error_Stability} shows the stability check for the proposed EKF model of D-FCTBN parameters. As shown in the Figure, the mean square error rapidly decreases with respect to the time and iterations, showing an acceptable level of stability for the analysis.

\Figure[ht!](topskip=0pt, botskip=0pt, midskip=0pt)[width=.9\textwidth]{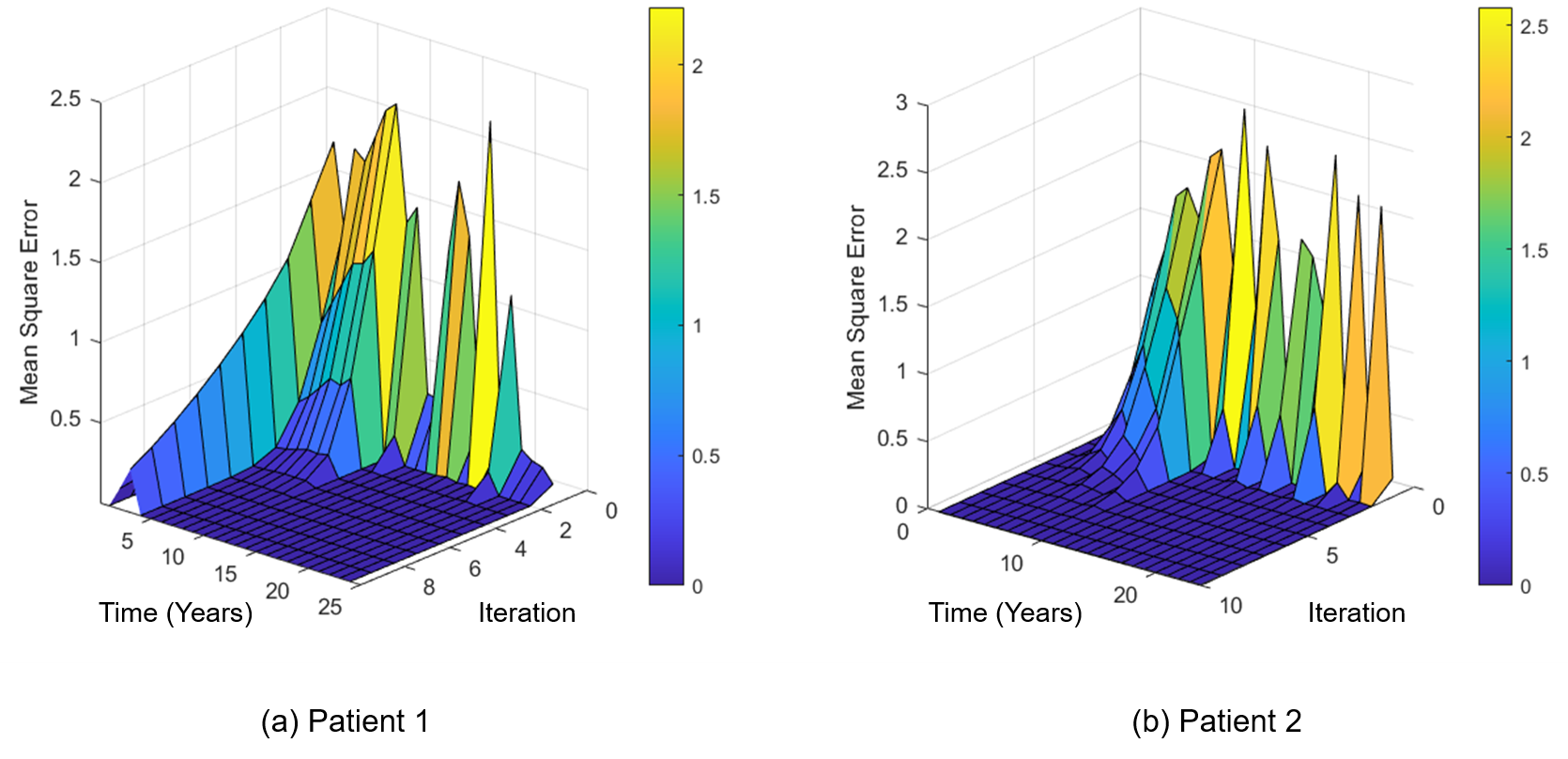}
{
The stability check of the the D-FCTBN model for estimating the model parameters in the presence of new data.
The Figure shows the MSE of predictions for two patients. The time axis shows the parameters estimated at each time step, and the iteration axis shows the steps to minimize the error at each time step.).\label{Figure:MSE_Error_Stability}
}


\subsection{Monitoring Events in a D-FCTBN}
\label{Subsection:Tensor_MPCA}
To demonstrate the effectiveness of the proposed tensor control chart for event detection for a change in D-FCTBN model parameter discussed in Section \ref{subsubsection: Monitoring_Scheme}, we setup two experiments, (1) a simulated experiment, where the patient's data and the behavioral data are synthetically generated to represent similar characteristics of the actual data, and (2) a real experiment, where we use the previously introduced data.
In the model setup, we have three non-modifiable demographic conditions (age, gender, and years of education) and four modifiable behavioral factors (healthy diet,  exercise, smoking habits, and drinking habits). We conducted the experiments in two stages. We utilize the data generated (the estimated $\beta_t$ coefficients) to build phase I of the control chart. During this period, the behavioral factors are controlled (for the simulated data). In phase II, out-of-control samples are randomly generated for the simulated data (by altering the modifiable factors). For each sample, the monitoring features and the residuals are calculated. The error features are then plotted on the corresponding control charts. 
To demonstrate the generative capability of the proposed approach, for simulated experiments, we consider/generate monthly data/observations. However, for the real experiments, we consider yearly time intervals, given the availability of data.
For the simulated experiments, two scenarios are considered. In the first scenario, we change only one of the behavioral factors, and in the second scenario, we change two behavioral factors simultaneously. Meanwhile, for real experiments, we consider the case where two behavioral factors change.

\Figure[t!](topskip=0pt, botskip=0pt, midskip=0pt)[width=0.9\textwidth]{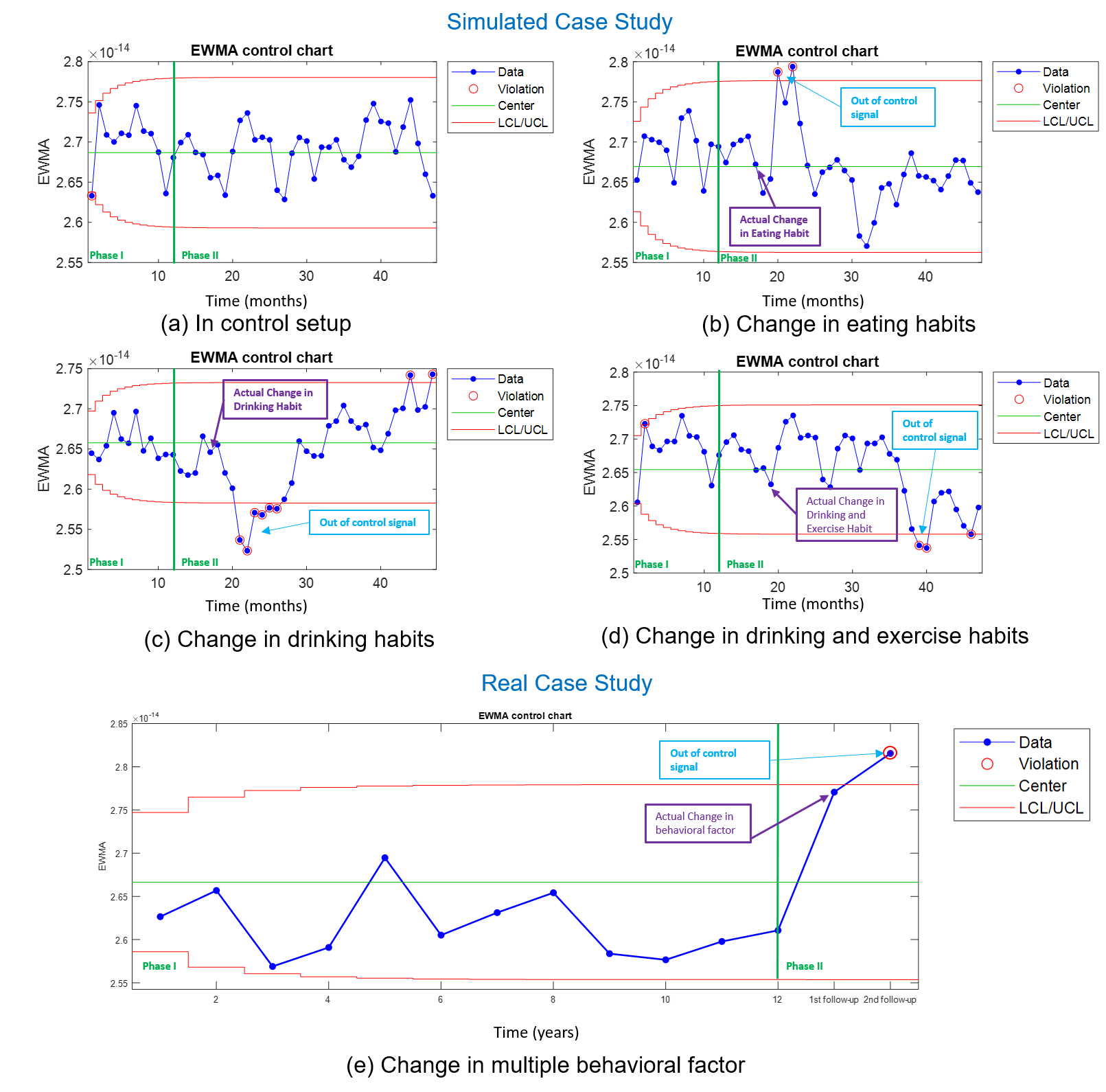}
{EWMA Control chart of the reconstruction error obtained from the proposed model. (a) In-control control chart (simulation case) (b-c) shows case 1 where only one of the behavioral factors are modified (simulation case), (d) shows case 2 where we randomly change more than one behavioral factor (simulation case), and (e) shows (uncontrolled) changes in more than one behavioral factor (real case)).\label{Figure:Control_Chart_Example}}

\subsubsection{Simulated Case Study: Changing One of the Behavioral Factors}
\label{Subsubsection:Case_1}
We consider the following setup for estimation of the parameters of the control chart (Phase I). The patient is considered to have diabetes as a prior chronic condition. The patient's fall is in the age range of 31-35 and male. The lifestyle behaviors during the in-control phase are [Healthy Diet, Exercise, Smoke, Drink] = [Yes, Yes, No, No]. We assume no extreme behavioral change during the phase I period (12 months). For evaluating the performance of the control chart (phase II) (after the first 12 months), we modify one or more of the behavioral factor/s of the patients at some (random) points in time. The control chart parameters are set to $\lambda$ = .15 and $L$ = 1.5 based on simulation analysis.

\noindent \textbf{Case (a): In-control behavior:} Figure \ref{Figure:Control_Chart_Example}\textcolor{blue}{(a)} shows the control chart for an in-control case where there is no behavioral change in either phase I or phase II (in 48 months). As a result, the control chart doesn't produce any out-of-control signal, which verifies its low type I error. 

\noindent \textbf{Case (b): Change in eating habits:} Figure \ref{Figure:Control_Chart_Example}\textcolor{blue}{(b)} represents the case, where the patient changes his diet from healthy eating to unhealthy eating, i.e. [Healthy Diet, Exercise, Smoke, Drink] = [No, Yes, No, No]. We introduce this change in the eating habit in the 17th month. The out-of-control events can be noticed in the control chart after three observations (in month 20). The quick diagnosis of the change in the behavioral change by the control chart can be attributed to the significant effect of eating habits on the parameters of the D-FCTBN.


\noindent \textbf{Case (c): Change in drinking habits:} For the next out-of-control scenario, we assume the patient picks up drinking alcoholic beverages i.e. [Healthy Diet, Exercise, Smoke, Drink] = [Yes, Yes, No, Yes]. This change was made on 17th month. 
The control chart picks up this change after four observations (around the 21st month). 
Similar to eating habits, the  quick detection of the change in the drinking habits by the control chart can be related to the significant effect of eating habits on the parameters of the D-FCTBN.

\subsubsection{Simulated Case Study: Changing More than One Behavioral Factors}
\label{Subsubsection:Case_2}
\noindent \textbf{Case (d): Change in drinking and exercise habits:} 
Here, we consider the same behavioral factors and set up as the phase I analysis. 
Meanwhile, for phase II analysis, we modify two behavioral factors (instead of one) simultaneously at on 19th month. 
The factors considered for change are [Healthy Diet, Exercise, Smoke, Drink] = [No, Yes, No, Yes].  
As shown in Figure \ref{Figure:Control_Chart_Example}\textcolor{blue}{(c)}, the first out-of-control signal is produced by the control chart in month 39 (after 20 months). This prolonged time for diagnosis can be because of the the complex interaction between the modifiable risk factors, which have changed in the opposite directions (stop exercise and stop drinking simultaneously).   

\subsubsection{Real Case Study: Change in Two Behavioral Factors}
\label{Subsubsection:Case_3}
For the real case study, we consider the patients' data presented in Section \ref{Subsection:EKF_Update}. Due to limited number of consecutive visits data available, we conducted this experiment in a hybrid setting. 
The patient considered has Hyperlipidemia and Obesity as a prior chronic condition/s. The lifestyle behaviors during the in-control phase are [Healthy Diet, Exercise, Smoke, Drink] = [No, Yes, No, Not Provided]. We estimate the data for 12 years using the initial learned model and build phase I control chart with this prior setup. Next, we utilize the patient actual behavioral factors in their follow-up meeting for phase II analysis.

\noindent \textbf{Case (e): Real behavioral change:} Figure \ref{Figure:Control_Chart_Example}\textcolor{blue}{(e)} shows the control chart with the real patient data in phase II. The figure shows the patient's behavior change in year 13 to [Healthy Diet, Exercise, Smoke, Drink] = [No, No, Yes, Not Provided]. Consequently, the chart produces an out-of-control signal after one observation (at year 14), which shows the sensitivity of the proposed control scheme when a significant behavioral change or multiple changes in the same (negative/positive) direction occurs.

\section{Conclusion}
\label{Section:Conclusion}
In this paper, we proposed extending functional continuous time Bayesian network based on an extended Kalman filter to build a dynamic functional continuous time Bayesian network to capture the dynamic (vs static) effect of changes of modifiable variables on the structure and parameters of the FCTBNs. We also utilized a low-rank tensor decomposition method based on multilinear principal component analysis to extract meaningful features of the tensor of D-FCTBN parameters. We developed a monitoring scheme based on the concept of a multivariate exponentially weighted moving average (MEWMA) control chart to identify changes in the modifiable variables that can significantly affect the structure and/or parameters of the D-FCTBN. We validated the proposed approach using both real data from Cameron County Hispanic Cohort (CCHC) as well as simulations. The results demonstrate the effectiveness of the proposed D-FCTBN and tensor control chart for dynamic prediction and monitoring the impact of patient's modifiable lifestyle behaviors on the emergence of multiple chronic conditions.



\bibliographystyle{IEEEtran}
\bibliography{EKF_ControlChart}

\EOD
\end{document}